\documentstyle[11pt]{article}
\begin{document}
\mbox{ }
\rightline{UCT-TP-209/94}
\rightline{March 1994}
\begin{center}
\begin{Large}
{\bf  Operator product expansion and duality at finite
temperature\footnote{Invited talk at {\it QCD-94}, Montpellier, July
1994}}
\end{Large}

{\large {\bf C. A. Dominguez\footnote{John Simon Guggenheim
Fellow 1994-1995}}}

Institute of Theoretical Physics and Astrophysics, University of
Cape Town, Rondebosch 7700, South Africa\\
\end{center}

\begin{abstract}
\noindent
The operator product expansion of current correlators at short distances,
and the notion of QCD-hadron duality are the cornerstone of QCD sum
rules. The extension of this programme to $T \neq 0$ is discussed,
together with applications to hot hadronic propagators. Indications
are that the hadronic spectrum suffers a substantial rearrangement
with increasing temperature, and hint on the existence of a quark
deconfining phase transition. Phenomenological order parameters to
characterize this phase transition are discussed.
\end{abstract}

\setlength{\baselineskip}{1\baselineskip}
\noindent

Some time ago Bochkarev and Shaposnikov \cite{ref1} proposed an extension of
the QCD sum rule program to non-zero temperature, and made an application
to the two-point function involving the vector current. This
application was reconsidered in \cite{ref2}.
Later on we discussed the axial-vector channel \cite{ref3} and the
nucleon channel \cite{ref4} using
Finite Energy QCD sum rules (FESR). Results from these analyses indicate
a substantial rearrangement of the hadronic spectrum with increasing
temperature, and hint on the existence of a deconfining phase transition.
This was later confirmed in \cite{ref5}, where a formalism valid even for
T near the critical temperature was used.
An extension of  QCD sum
rules to finite temperature entails the assumptions that (a) the OPE
continues to be valid, except that now the vacuum condensates will develop
an ({\it a-priori}) unknown temperature dependence, and (b) the notion of
QCD-hadron duality also remains valid. At T=0,
the thermal behaviour of the vacuum condensates is not calculable
analytically from first
principles; some model or approximation must be invoked, e.g. the dilute
pion gas approximation, lattice QCD, etc..
At $T=0$, QCD-hadron duality is well established
phenomenologically, while the validity of the OPE in QCD (in the presence of
non-perturbative phenomena) is made plausible after studying other
field theories which can be solved exactly \cite{ref6}.
At $T\neq 0$, it is important to examine critically the basis for the
two assumptions above, especially in the light of some recent adverse
criticisms \cite{ref7}.\\
The basic object under consideration is the retarded (advanced) two-point
function, involving a local current operator $J(x)$,
after appropriate Gibbs averaging
\begin{equation}
\Pi \; (q,T) = i \int d^{4}x \;\exp (i q x) \; \theta (x_{0})\;
 << [J(x), J^{\dag}(0)] >> \; ,
\end{equation}
where
\begin{equation}
<< A \cdot B >> = \sum_{n}\; \exp (-E_{n}/T)\; \langle n|A \cdot B|n \rangle
\;/Tr (\exp (-H/T)) \; ,
\end{equation}
and $\mid n >$ is a complete set of eigenstates of the (QCD) Hamiltonian.
The OPE of $\Pi (q,T)$ is formally written as
\begin{equation}
\Pi \; (q,T) = C_{I} << I >> + \sum_{r} C_{r} (q) << {\cal{O}}_{r} >> \; ,
\end{equation}
The unit operator $I$ in Eq.(3) represents the purely perturbative piece,
and the
OPE  is assumed valid, even in the presence of non-perturbative effects,
for $q^{2} < 0$ (spacelike), and  $\mid q^{2} \mid $ $\gg \Lambda_{QCD}^{2}$.
The states $\mid n>$ entering Eq.(2) can be any complete set of states, e.g.
hadronic states, quark-gluon basis, etc.. If one wishes to extend smoothly
the $T=0$ QCD sum rule program  to finite temperature, then the natural
choice for this set is the quark-gluon basis, as first proposed
in \cite{ref1}. I discuss below some arguments in support
of the validity of the OPE at
$T \neq 0$, Eq.(3). Assuming this, and invoking duality at $T=0$, one now
raises the temperature by an arbitrary small value. The hadronic spectrum
and the expectation values of the QCD operators in the OPE
will hardly change. Hence, it is reasonable to expect that the
inter-relationship between QCD and hadronic parameters effected by
duality will remain valid. An abrupt dissapearance
of this inter-relationship as soon as $T \neq 0$, as advocated in
\cite{ref7}, appears as an unlikely possibility. One
should keep in mind that the current in Eq.(1) is an object external to
the heat bath, and thus need not be in thermal equilibrium with the
particles in the medium. The fact that at low $T$
these particles are predominantly hadrons, is irrelevant for the argument.
The external current will still convert to quark-antiquark pairs, which
contribute e.g. to the perturbative operator $I$ through loops, where
they can have any value of momentum. In addition, there will be thermal
non-perturbative effects as parametrized by the condensates. These are
always distinguishable from perturbative effects, since they have a
different $q^{2}$ dependence. In many cases they have also a different
T-dependence, and may contribute only to the real part of the
correlators. As the temperature is increased
further, the hadronic spectrum will begin to change shape, in pace with
the thermal behaviour of the expectation values of the various
operators in the OPE. I shall discuss later how these changes
emerge from  QCD sum rules in agreement with expectations from
general physical considerations. \\
I turn now to the validity of the OPE at $T \neq 0$. Just as at $T=0$, no
rigorous proof can be given since one cannot solve QCD analytically and
exactly. Instead, other field theory models which can be solved exactly
have been used to argue for the validity of the OPE at $T=0$ \cite{ref6}.
These models include the $O(N)$ sigma model in the large $N$ limit, and
the Schwinger model, both in two dimensions. Comparing the short distance
expansion of the exact solution for a Green function with the OPE counterpart,
one finds that they are identical \cite{ref6}. One can also show that
this is also the case at finite temperature \cite{ref8}, as I
outline next.
Let us consider first the O(N) sigma model in 1+1 dimensions which is
characterized by the Lagrangian
\begin{equation}
{\cal{L}} = \frac{1}{2} [\partial_{\mu} \; \sigma^{a} (x)] \;
[\partial_{\mu} \sigma^{a} (x)] \; ,
\end{equation}
where a = 1,...N and $\sigma^{a} \sigma^{a} = N/f$, with f being the coupling
constant. In the large N limit this model can be solved exactly (for
details see \cite{ref6}), it is known to be asymptotically free,
and in spite of the absence of mass parameters in Eq.(4), it exhibits
dynamical mass generation. In addition, in this model there are vacuum
condensates, e.g. to leading order in 1/N :
$\langle 0| \alpha |0 \rangle = \sqrt{N} \; m^{2} \;$ ,
whereas all other condensates factorize, viz.
$\langle 0| \alpha^{k} |0 \rangle = (\sqrt{N} \; m^{2})^{k} \;$ .
The $\alpha$ field is: $\alpha = f (\partial_{\mu}
\sigma ^{a})^{2}/\sqrt{N}$, and  we are interested in the Green function
associated with the propagation of quanta of this $\alpha$ field.
We have calculated this Green function  at finite temperature. Its
imaginary part can be integrated analytically in closed form and is
\begin{eqnarray}
\mbox{Im} \; \Gamma (\omega, {\bf q} = 0, T) &=&
\frac{1}{2 \omega^{2}} \left[ 1 + 3 n_{B} \; (\omega/2T) \right] \nonumber
\\[.4cm]
& + & \frac{1}{2} \left[ \frac{2}{\sqrt{N}} \;
\frac{<< \alpha >>}{\omega^{4}} + \frac{6}{N} \;
\frac{<< \alpha^{2} >>}{\omega^{6}} \;  + \; \cdots \right] \; ,
\end{eqnarray}
where the first term above corresponds to the perturbative contribution,
the second to the non-perturbative, and $n_{B}$ is the thermal Bose factor.
Equation (5) is valid in the time-like region; the space-like region
counterpart vanishes in 2 dimensions.
Since the model is exactly solvable, the thermal behaviour of the vacuum
condensates can also be calculated, viz.
\begin{equation}
<< \alpha >> = < \alpha > \left[ 1 + 3 n_{B} (\omega/2T)
\right]
\end{equation}
In this case the vacuum condensates contribute to the imaginary part, and
as Eq.(5) shows, the thermal dependence of the
perturbative piece cannot be absorbed into the condensates. Hence, no
confusion should arise between perturbative and non-perturbative
contributions.\\
Finally, I consider the Schwinger model in 1+1 dimensions, with the
Lagrangian
\begin{equation}
{\cal{L}} = - \frac{1}{4} F_{\mu \nu} \;
F_{\mu \nu} + \bar{\psi} i \;
\gamma_{\mu} \; {\cal{D}}_{\mu} \; \psi
\end{equation}
where ${\cal D}_{\mu} = i \partial_{\mu} + e A_{\mu}$. This model has been
solved exactly, and in the framework of the OPE \cite{ref6}.
The short distance expansion of the exact solution coincides with that
from the OPE \cite{ref6}.
Here, we are interested in the two-point functions
\begin{equation}
\Pi_{++} (x) = \langle 0 |T \{j^{+} (x) j^{+} (0) \}|0 \rangle
\end{equation}
\begin{equation}
\Pi_{+-} (x) = \langle 0 |T \{j^{+} (x) j^{-} (0) \}|0 \rangle
\end{equation}
where the scalar currents are:
$j^{+} = \bar{\psi}_{R} \; \psi_{L} \;\;$,
$j^{-} = \bar{\psi}_{L} \; \psi_{R}\;\;$,
with $\psi_{L,R} = (1 \pm \gamma_{5}) \psi/2$. The function $\Pi_{++} (Q)$
vanishes identically in perturbation theory, and the leading non-perturbative
contribution involves a four-fermion vacuum condensate.
We have calculated the thermal behaviour of these  current
correlators  and obtain, e.g. for their imaginary parts in the
time-like region (again, there is no space-like contribution in 2 dimensions)
\begin{equation}
\mbox{Im} \; \Pi_{++} (\omega, {\bf q} = 0,T) = 0
\end{equation}
\begin{equation}
\mbox{Im} \; \Pi_{+-} (\omega, {\bf q} = 0,T) = \frac{1}{4} \;
\left[ 1 - 2n_{F} (\omega/2T) \right]
\end{equation}
Hence, the choice of the fermion basis in the Gibbs average of current
correlators does not imply confusing these fermions with condensates, as
argued in \cite{ref7}. As
Eqs.(10) and (11) indicate, (perturbative) fermion loop terms and
(non-perturbative) vacuum condensates develop their own temperature
dependence, which in this particular example happen to be different.\\
Turning to applications, an inspection of a typical hadronic spectral
function at $T=0$ shows some resonance peaks at low energy, followed
by a hadronic continuum starting at some threshold $s_{0}$. Because
of asymptotic freedom, this continuum is well approximated
by perturbative QCD. Raising the temperature, one would expect resonance
melting. This may be accomplished by  the imaginary part of hadronic
propagators growing with temperature, as first proposed in \cite{ref9}.
As the resonance peaks become broader, one would expect $s_{0}$ to
decrease. Close to the critical temperature for deconfinement, the
hadronic spectral function should be rather smooth, and represented almost
entirely by the quark-gluon degrees of freedom over the whole energy
range. An indication for some decrease of $s_{0}$ with increasing $T$
was obtained from QCD sum rules for the vector channel \cite{ref1}
-\cite{ref2}, \cite{ref10}. However, it is the axial channel the one
that provides conclusive evidence for this behaviour \cite{ref3},
\cite{ref5}. In fact, the solution to the FESR gives
\begin{equation}
\frac{s_{0}(T)}{s_{0}(0)}  \simeq  \frac{f_{\pi}^{2}(T)}
{f_{\pi}^{2}(0)}\; .
\end{equation}
The temperature behaviour of the imaginary part of
a hadronic propagator is difficult to obtain
from QCD sum rules, since these are not too sensitive to resonance widths.
Independent calculations in other theoretical frameworks
\cite{ref11} show that the
pion, the nucleon, and the rho-meson propagators have imaginary parts
which grow monotonically with $T$. Over a certain range of temperatures
these behave as
\begin{equation}
\Gamma (T) = \frac {\Gamma (0)}{(1 - T/T_{c})^{\alpha}}\;.
\end{equation}
Thus, the asymptotic freedom threshold $s_{0}(T)$, and the resonance
width $\Gamma (T)$, are suitable phenomenological order parameters to
characterize the quark deconfinement phase transition.\\
It should be mentioned, in closing, that there are some unresolved
problems in finite temperature QCD sum rules, e.g. when one tries
to use them to extract the thermal behaviour of the vacuum condensates
\cite{ref10}. An unwelcome implication is a breakdown of the FESR
program beyond the lowest moment sum rule, and hence a breakdown of
Laplace transform QCD sum rules. A satisfactory resolution of this
problem is eagerly awaited.
\subsection*{Acknowledgements}
This work was done in collaboration with M. Loewe. The author wishes to
thank Stephan Narison for a perfectly organized, and most enjoyable
workshop.

\end{document}